\documentclass{jfm}
\usepackage{graphicx}
\usepackage{epstopdf, epsfig}
\usepackage{amsmath}
\usepackage{amsbsy}

\usepackage[usenames,dvipsnames,svgnames,table]{xcolor}
\usepackage{color,soul}
\usepackage{color}
\shorttitle{A linearized model}
\shortauthor{M. Garcia, B. Ganapathysubramanian and S. Pennathur}

\title{A linearised model for calculating inertial forces on a particle in the presence of a permeate flow}

\author{Mike Garcia\aff{1}
  \corresp{\email{mikegarcia@ucsb.edu}},
  B. Ganapathysubramanian\aff{2}
 \and S. Pennathur\aff{1}}

\affiliation{\aff{1}Department of Mechanical Engineering, University of California Santa Barbara,
Engineering II, Room 2355 University of California, Santa Barbara, CA 93106

\aff{2}Department of  Mechanical Engineering, Iowa State University, 306 Lab of Mechanics, Ames, IA 50011}

\begin{document}
\maketitle
\begin{abstract}
Understanding particle transport and localisation in porous channels, especially at moderate Reynolds numbers, is relevant for many applications ranging from water reclamation to biological studies. Recently, researchers experimentally demonstrated that the interplay between axial and permeate flow in a porous microchannel results in a wide range of focussing positions of finite sized particles \citep{Garcia:2017ho}. We numerically explore this interplay by computing the lateral forces on a neutrally buoyant spherical particle that is subject to both inertial and permeate forces over a range of experimentally relevant particle sizes and channel Reynolds numbers ($\Rey$). Interestingly, we show that the lateral forces on the particle are well represented using a linearised model across a range of permeate-to-axial flow rate ratios, $\gamma$. Specifically, our model linearises the effects of the permeate flow, which suggests that the interplay between axial and permeate flow on the lateral force on a particle can be represented as a superposition between the lateral (inertial) forces in pure axial flow and the viscous forces in pure permeate flow. We experimentally validate this observation for a range of flow conditions. The linearised behaviour observed significantly reduces the complexity and time required to predict the migration of inertial particles in permeate channels. 
\end{abstract}

\begin{keywords}
Authors should not enter keywords on the manuscript, as these must be chosen by the author during the online submission process and will then be added during the typesetting process (see http://journals.cambridge.org/data/\linebreak[3]relatedlink/jfm-\linebreak[3]keywords.pdf for the full list)
\\
\end{keywords}


\section{Introduction}
Lateral migration and focusing of neutrally buoyant solid particles at moderate Reynolds number ($\Rey$) in a confined pressure driven flow is a well known phenomenon first documented by Segr\'e and Silberberg in 1961. Specifically, hydrodynamic inertial stresses cause particles to laterally migrate across streamlines and ultimately focus into distinct locations in the channel \citep{Ho:1974eh}. In addition to the seminal work on inertial migration and focusing \citep{Cox:1968ce,Ho:1974eh,Saffman:1965eo} there have also been a few recent reviews highlighting progress \citep{Martel:2014et, Amini:2014cm, Zhang:2016dp}. However, comparatively, there have only been a few studies on the motion of inertial particles in the presence of a permeate flow. Systems with such flows are ubiquitous in applications related to pressure-driven membrane filtration for the separation of particles and cells from liquid suspensions as well as in a multitude of other areas, including wastewater treatment \citep{Chang:2005co}, food \citep{FernandezGarcia:2013jd} and beverage \citep{Pall2015} processing,  and biotechnology \citep{Palmer:2009gr, Charcosset:2006wo}. 

In general, to precisely solve for the forces on a particle in inertial migration problems, one relies on simulation of the combined fluid-particle interaction problem. For example, Chun and Ladd \citep{Chun:2006eq} used the lattice-Boltzmann method for $\Rey$ ranging from 100 to 1000 to show that spherical particles migrate to one of a finite number of equilibrium locations in the cross-section of a non-porous square channel, where only the face centred equilibrium locations are stable for lower $\Rey$. Di Carlo \textit{et al.} \citep{DiCarlo:2009ee} performed simulations to show how particle equilibrium locations varies with particle sizes (a trend not captured by previous asymptotic approaches). 
Similarly, Liu \textit{et al.} \citep{Liu:2015ht} performed exhaustive simulations to determine the non-trivial relationship between the focusing location of particles and their size, channel aspect ratio, and $\Rey$. These numerical studies -- fit into generalised formulae -- shed light on how the forces on a migrating particle are distributed within a channel and how this distribution depends on particle size. Such findings are crucial for practical device design.

With the addition of permeate flow, the interplay between the effects of axial and permeate flow (see figure 1a) results in a much richer particle behaviour than the case of pure inertial migration of a particle in axial flow \citep{Altena:1984ft,Drew:1991iq,Lebedeva:2011kj}. Earlier theoretical work by Belfort and coworkers accounts for the effect of wall porosity on particle motion in a 2D geometry \citep{Altena:1984ft,Drew:1991iq} and has been validated experimentally \citep{1986ExFl....4....1O}. In these theoretical studies, researchers employed asymptotic analysis to derive expressions for the forces on a migrating particle as a sum of inertially induced forces and permeate drag. While useful from a theoretical perspective, the researchers were unable to precisely predict the lateral lift forces of highly confined particles in a porous channel. This limitation is particularly critical since precise force values across the cross-section are required for efficient device design that exploit the lateral migration of particles to separate and concentrate particles with high specificity.  More recently, Garcia and Pennathur \citep{Garcia:2017ho} experimentally demonstrated that a permeate flow can drastically alter the migration and focusing of confined particles in inertial flow, with particle size significantly impacting resulting migration behaviour. However, no model to date has accurately described this behaviour. The ability to construct a simple model that reliably predicts these forces across a wide range of conditions serves as the overarching motivation for the present work.

Based on observations from a finite number of full-scale simulations, we develop a linear model to efficiently predict the lateral forces acting on a neutrally buoyant spherical particle migrating in a pressure driven porous channel. We first employ a previously utilised full-physics simulation approach \citep{DiCarlo:2009ee, Kim:2016jv, Gossett:2012bx, Hood:2015iz} to numerically simulate a spherical particle translating in a square cross-section microchannel. We additionally validate this model against previously published work and experiments, then exploit the model to study the lateral forces in a system with permeate flow of varying magnitude and direction. We observe that the relative permeate flow, $\gamma$, can generalise the data over many particle sizes and $\Rey$, which provides a conceptual basis for developing a new linear model. We finally validate our linear model using the full-physics simulation, and determine that the limits of applicability are well within those of most experimental systems. To prove as much, we design a microfluidic system to experimentally validate the results of the linear model. 


\section{Full Physics Simulations of Particle-Fluid Interaction} 
The present work focuses on the combined inertial and permeate migration of a single neutrally buoyant spherical particle in Poiseuille flow within the confining geometry of a porous channel of square cross-section ($W \times W$) and length \textit{$L_C$ = 10W}. Here the particle of diametre $a$ is translating with velocity $\textbf{U}_P=[0, U_P, 0]$ and angular velocity $\boldsymbol{\Omega}$ in a flow of average (local) velocity $U = (U_{in}+U_{out})/2$, where $U_{in}$ and $U_{out}$ are the average axial flow velocities at the inlet and outlet of the porous domain respectively. The porous channel allows for a lateral flow to exist that emanates from only two parallel walls at a constant velocity $U_W$. Here we define the channel Reynolds number $\Rey = \rho U W / \mu$ and the relative permeate flow $\gamma = U_{W}/U$, where $\rho$ and $\mu$ are the fluid density and viscosity respectively. 

\begin{figure}
\centering
\centerline{\includegraphics{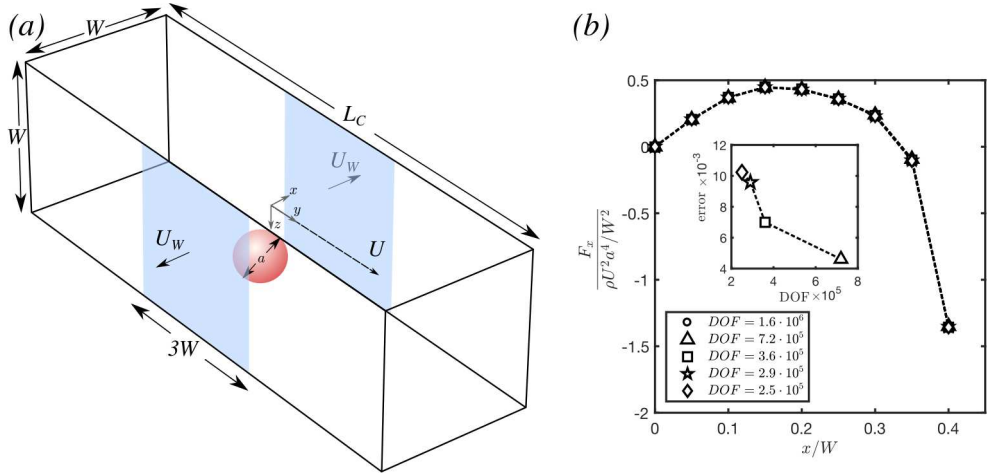}}
\caption{ (\textit{a}) Schematic illustration of our system. We model a square channel with average velocity $U$. In the channel a spherical particle of diametre $a$ is migrating within the confines of the bounding walls where two of the walls are permeable ($yz$-plane) and allow flow to penetrate at a constant rate of $U_W$. For each $x-z$ location of the particle in the channel cross section, the lateral lift forces ($F_z$ and $F_x$) are calculated.
(\textit{b}) Mesh sensitivity analysis showing that the calculated lift forces have converged and are, thus, insensitive to the degrees of freedom (DOF) in the model. The inset depicts the error relative to a model with $1.6 \times 10^6$ DOF. ($\Rey =100$, $a/W = 0.1$).}
\label{fig:F1}
\end{figure}

We consider a moving frame of reference translating with the accelerating particle \footnote{ In this moving reference frame, the lateral flow extends a length $3W$ as depicted in figure \ref{fig:F1}}. In this reference frame, we solve the steady Navier-Stokes equations: 

\begin{equation} \label{EQ1}
\rho\big( \mathbf{u} \cdot \nabla \mathbf{u}  + \textbf{U}_{P} \cdot \nabla \bar{\textbf{u}}\big) = \mu \nabla^2 \mathbf{u} - \nabla p \\
\end{equation}

\begin{equation} \label{EQ2}
   \nabla \cdot  \textbf{u} = 0
\end{equation}
where $p$ is the fluid pressure, $\textbf{u}$ is the fluid velocity in the reference frame of the translating particle and $\bar{\textbf{u}}$ is the undisturbed flow. Note that the second inertial term arises due to the acceleration of our chosen reference frame as shown in more detail in (appendix \ref{App_A}).

The velocity of the suspended particle ($U_P$ and $\boldsymbol{\Omega}$) can be self-consistently determined by setting conditions such that the axial motion satisfies a drag constraint $F_{y} = 0$ (equation \ref{EQ3}) and its rotational motion satisfies a torque constraint $\tau_{x} = \tau_{y} = \tau_{z} = 0$ (equations \ref{EQ4}).

\begin{equation} \label{EQ3}
F_{y} \equiv       \textbf{e}_y \cdot   \int_s  \textbf{n\textsubscript{r}} \cdot \textbf{T} \,ds  = 0 
\end{equation}

\begin{equation} \label{EQ4}
\tau_i \equiv  \textbf{e}_i \cdot \int_s  (\textbf{r}-\textbf{r}_{p})  \times  \textbf{n\textsubscript{r}}  \cdot \textbf{T} \,ds= 0\quad \textrm{where $i = x,y,z$}
\end{equation}

where the integrals are over the surface of the sphere, $\textbf{r}_{p}$ is the particle position vector that points from the centre of the channel to the centre of the particle and  $\textbf{n\textsubscript{r}}   = (\textbf{r}-\textbf{r}_{p}) / |\textbf{r}-\textbf{r}_{p}|$ is the unit normal at each point on the surface of the sphere. $\textbf{T}$ is the total stress tensor and, for an incompressible Newtonian fluid, is given by  $\textbf{T} = -p \mathsfbi{I} + 2 \mu \boldsymbol{E}$, where $\textbf{E} = \frac{1}{2}(\nabla \mathbf{u} + \nabla \mathbf{u} ^{T})$ is the rate of strain tensor. 

The boundary conditions of this problem are in the reference frame of the translating particle. Therefore, the no slip condition on the wall is:
\begin{equation} \label{EQ5}
\textbf{u}  = -U_{p} \textbf{e}_y  \quad  \textrm{on all walls}
\end{equation}
additionally, the channel has two parallel porous walls that can advect fluid into or out of the channel at a constant rate of $U_W$. Thus the condition on these walls also must satisfy:
\begin{equation} \label{EQ6}
\textbf{u} \cdot \textbf{n} = U_{W}
\end{equation}
where $\textbf{n}$ is the wall unit normal pointing out of the channel and  a positive value of $U_{W}$ indicates flow out of the channel. The no slip condition on the rotating particle is enforced by assigning a velocity to the surface of the sphere corresponding to that of rigid body rotation at angular velocity, $\boldsymbol{\Omega}$:
\begin{equation} \label{BC2}
\textbf{u}_{surface} = \boldsymbol{\Omega} \times (\textbf{r}-\textbf{r}_{p}) 
\end{equation}
Finally, far from the particle the flow field satisfies

\begin{equation} \label{BC3}
\textbf{u} =  \bar{\textbf{u}} - U_{p}\textbf{e}_y
\end{equation}

To solve for the unknowns (i.e., $\mathbf{u}$, $p$, $U_{p}$ and $\boldsymbol{\Omega}$) we couple equation \ref{EQ3} and equation \ref{EQ4} to the fluid flow equations and solve directly using the COMSOL multiphysics software. This procedure is performed for a lattice of discrete positions of the particle within the cross-section of the channel (using only a quarter of the domain via symmetry arguments to minimise computational effort). To calculate the lift force on the particle, we integrate the surface stresses on the particle in the appropriate direction ($x$ or $z$):

\begin{equation} \label{EQ8}
F_{i} = \textbf{e}_i \cdot   \int_s    \textbf{n}_r  \cdot  \textbf{T} \,ds \quad \textrm{where $i = x,z$}
\end{equation}


\section{Numerical Results}\label{sec:NR}

\subsection{Mesh convergence analysis}

We discretise the fluid domain using tetrahedral elements in COMSOL Multiphysics. The surface of the sphere is discretised into triangular boundary elements (BE) with 6 boundary layer (BL) elements at the surface of the sphere to accommodate large gradients. We use quadratic basis functions for representing the velocity solution and linear basis functions for the pressure. The resulting discretisation has approximately $3.6 \times 10^{5}$ degrees of freedom (DOF). To show that the calculated results were independent of mesh density we increased the number of DOF in our model by varying the BE, BL and fluid domain tetrahedral density. The resulting calculated lift forces are shown in figure 1b, and show no significant differences ($\leq 1\%$ change) in the spatial lift force amongst the different cases.

\begin{figure}
\centering
\centerline{\includegraphics{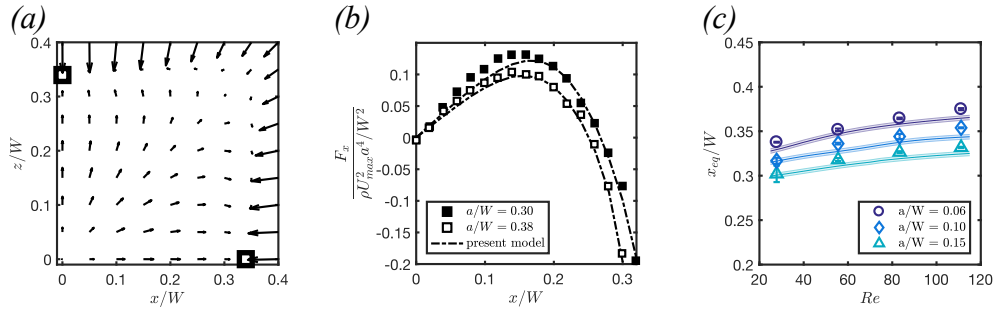}}
\caption{(\textit{a}) Inertial forces calculated in a single quadrant of the square cross-section for a particle of $a/W = 0.10$ at $\Rey =100$. Here the black small squares indicate the locations of the stable equilibrium points, \textit{i.e.} locations where the lift forces go to zero. (\textit{b}) A comparison of the $x$-component inertial forces along the $z/W = 0$ axis between the present study and that of Di Carlo {\it et al} \citep{DiCarlo:2009ee}, showing good agreement at $\Rey = 38$ for two particles $a/W = 0.30$ and $a/W = 0.38$. (\textit{c}) Experimental measurements (discrete points) from Garcia and Pennathur \citep{Garcia:2017ho} of the inertial focusing locations ($x_{eq}/W$) as a function of $\Rey$ for three particle sizes ($a/W$) with overlaid corresponding numerical simulations (solid lines).}
\label{fig:F2}
\end{figure}

\subsection{Model validation}
To demonstrate validity of the model, we performed simulations of a particle within the channel with no permeate flow ($U_W = 0$), since this reduces to the well-studied case of an impermeable channel \citep{DiCarlo:2009ee,Miura:2014ig,Shichi:2017bg}. Typically in experimentally relevant flows for inertial microfluidic systems, particle diametres are a significant fraction of the channel width ($a/W \geq 0.10$) and  generally $\Rey$ ranges from $\sim 10$ to $100$. In the case of an impermeable channel under these conditions, particles focus to four symmetric equilibrium positions near the centre of each wall face and approximately $0.3W$ away from the channel centre \citep{DiCarlo:2009ee}. Figure 2a provides a detailed map of the spatially varying inertial lift forces in a single quadrant of a channel. With this map, we can identify the location of zero lateral lift force, which agrees with previous experimental studies \citep{Miura:2014ig,Shichi:2017bg}. In figure 2b, we compare the spatial variation of the lift forces on the particle which show good agreement with data from Di Carlo \textit{et al.} (figure \ref{fig:F2}b). Finally, we show that our model can correctly predict changes in the locations of equilibrium with increasing $\Rey$ \citep{ASMOLOV:1999ef} and increasing particle size \citep{DiCarlo:2009ee} in figure 2c, where we compare our previous experimental data \citep{Garcia:2017ho}, with current simulation results. 

 \subsection{Permeate flow results}\label{sec:NumRes}
 
\begin{figure}
\centering
\centerline{\includegraphics{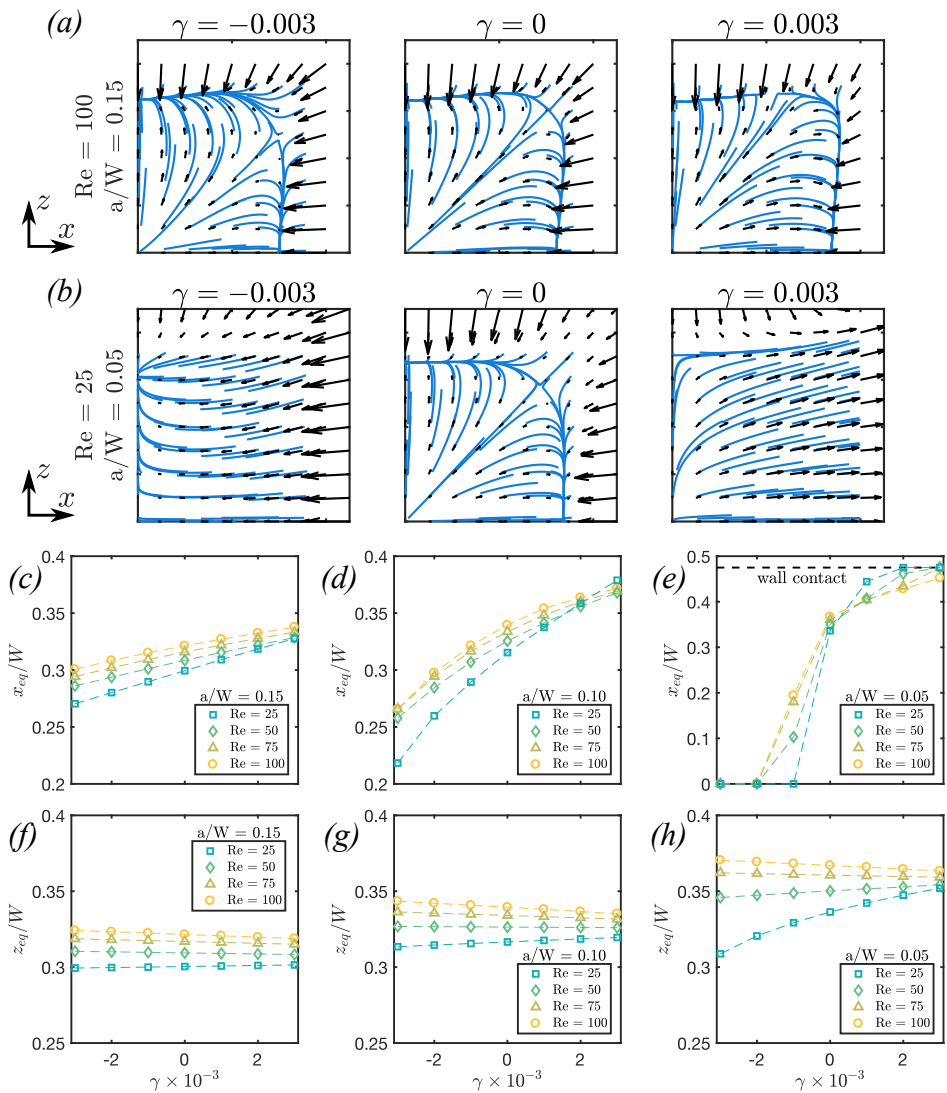}}
\caption{Lateral force vector fields spanning the extremities of the parameter space $\gamma = [-0.003:0.001:0.003]$ and $\Rey = [25:25:100]$ for two particle sizes (\textit{a}) $a/W = 0.15$ (\textit{b})  and $a/W = 0.05$. The blue `streamlines' are for visualisation of the vector fields which are bounded by a $0.45W$ $\times$ $0.45W$ box. (\textit{c-e}) $x$-equilibrium locations for all direct simulations in this study along the $z/W = 0$ axis for (\textit{c}) $a/W = 0.15$ (\textit{d}) $a/W = 0.10$ and (\textit{e}) $a/W = 0.05$. Note that the equilibrium shift is in the direction of the permeate flow. (\textit{f-h}) $z$-equilibrium locations for all direct simulations in this study along the $x/W = 0$ axis for (\textit{f}) $a/W = 0.15$ (\textit{g}) $a/W = 0.10$ and (\textit{h}) $a/W = 0.05$.}
\label{fig:F3}
\end{figure}

We next consider the effects of permeate flow, for a range of $\Rey$, $\gamma$ ($\gamma = U_W/U$), and particle sizes. We consider $\Rey = 25, 50, 75, 100$,  $\gamma = \pm 0.003, \pm 0.002, \pm 0.001, 0$ and particle sizes $a/W = 0.05, 0.1, 0.15$. Figure 3 a \& b plots the lateral lift force vectors for a subset of the simulations. 
Figures \ref{fig:F3}c-h show the equilibrium locations for the full range of $\gamma$ and $\Rey$, with figures \ref{fig:F3}a \& b showing full force vector fields for the extreme cases. At large Reynolds number ($\Rey$) and large particle size ($a/W$), the vector fields are only slightly disturbed by the presence of the permeate flow when compared to the case with no permeate flow ($\gamma = 0$) (\textit{e.g.} figure \ref{fig:F3}a), presumably because inertial lift forces are more dominant (since the lift force $F_L \sim a^4$ \citep{ASMOLOV:1999ef}). Conversely, for flows with less inertia (\textit{i.e.} smaller $\Rey$ and $a/W$) (\textit{e.g.} figure \ref{fig:F3}b) the vector fields are highly disturbed due to the substantial effects of permeate flow. Specifically, for $\gamma = 0.003$ (\textit{e.g.} figure \ref{fig:F3}b right), we observe a complete suction where the force equilibrium coincides with the location where wall contact occurs and is when all `streamlines' appear to be guided towards the porous wall ($x/W = 0.5$)\footnote{We do not actually model the physics of wall contact, when referring to wall contact we do so in the context of the particle equilibrium location \textit{i.e} $x_{eq}/W \geq 0.5-a/2W$.}. Conversely, when $\gamma = -0.003$ (\textit{e.g.} figure \ref{fig:F3}b left), there is a reversal of 'streamlines', now directed towards the centreline ($x/W = 0$).

Figures \ref{fig:F3}c-h captures all these observations as we plot the $x-$ and $z-$ force equilibrium location ($x_{eq}/W$ and $z_{eq}/W$) for all $\Rey$ and $a/W$. Through this representation we can clearly see that as permeate flow increases, the equilibrium location of each vector field shifts in the direction of the permeate flow (figures \ref{fig:F3}c-e). Furthermore, in agreement with the qualitative observation in figure \ref{fig:F3}a,b the shift in equilibrium location becomes less sensitive with increasing $a/W$ and $\Rey$. We also see a slight shift in the $z$-equilibrium with the addition of a permeate flow, but in general the $z$-equilibrium shift is less sensitive than the $x$-equilibrium counterpart. One thing to note is that the $z$-equilibrium shifts either towards or away from the wall as a function of $\gamma$, depending on the flow \Rey.In general, there is a clear dependence on two factors, the first being the magnitude of permeate forces (set by $U_W$) and the second being the magnitude of inertial forces (set by $U$). The combination of these two factors dictates the behavior of the particles in our geometry, ultimately leading to non-trivial force fields.

  \begin{figure}
\centering
\centerline{\includegraphics{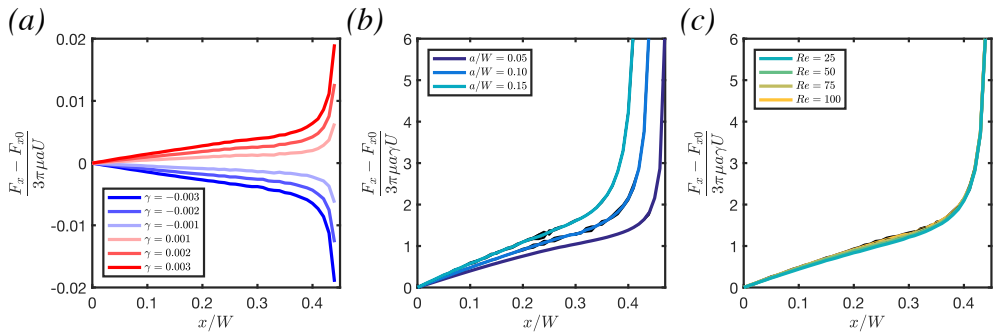}}
\caption{(\textit{a}) Residual force plot. Each curve represents a different value of $\gamma$, where the residual is the difference in the force between a particle in the presence of permeate flow and a particle in a channel with no permeate flow ($\gamma = 0$) at different $x$ locations. (\textit{b}) The force residuals for all $\gamma$ normalised by a characteristic drag force resulting in three distinct master curves corresponding to each particle size at $\Rey =100$. (\textit{c}) Normalised force residuals for four distinct $\Rey$ at a constant particle size ($a/W = 0.10$). Note that normalising in this manner results in the collapse of the curves into a single `master curve'. Due to noise in our simulations, we averaged values in both (b) and (c) as shown in the colored lines. Black curves underneath represent the raw normalisations).}
\label{fig:F4}
\end{figure}
 
\subsection{Linearised Model}
Solving for the effect of combined inertial and permeate forces with full-physics numerical simulations can quickly become computationally infeasible. For instance, the results in the previous sub-section involved performing $4(\Rey) \times 3(a/W) \times 9 (x/W) \times 9 (z/W) \times 7 (\gamma) = 6804$ simulations, that required expending substantial computational resources. The computational requirements become even larger when modeling dynamic processes where a particle may be migrating in a continuously varying flow field, or in which $\gamma$ is spatially and/or temporally changing. Therefore, we develop a linearised model (based on observations from the previous sub-section) to produce quantitatively similar results to the full simulation, but with significantly less computational requirements.

We consider a linearisation of the lateral lift force $\textbf{F}(\gamma, \Rey, a/W, x/W, z/W)$ about the $\gamma = 0$, \textit{i.e.} impermeable channel case. There are multiple reasons for choosing such a linearisation strategy: (a) This builds upon available lift force results for the impermeable channel case, which enables easy generalisation to other cross-sections, (b) this takes advantage of the fact that parameter $\gamma$ is naturally very small for the purposes of linearisation (since $|\gamma| \leq 0.01$ in typical microfiltration processes \citep{Belfort:1994if}), and (c) such a linearisation has prior analytical precedent that suggests an additive decomposition of the lift force into an inertial lift force and a permeate drag force \citep{Altena:1984ft}. 

We write the linearisation as 
\begin{multline} \label{EQ9}
\textbf{F}(\gamma, \Rey, a/W, x/W, z/W) \sim  \textbf{F}_{0}(\Rey, a/W, x/W, z/W)|_{\gamma = 0} 
\\ + \gamma \textbf{F}_{1}(\Rey, a/W, x/W, z/W)  + \mathcal{O}(\gamma^2)   
\end{multline}

where $\textbf{F}_{0}$ is the lift force calculated for a particle at a given location in the absence of any permeate flow (\textit{i.e.}, $\gamma = 0$, which is the standard inertial migration in an impermeable channel) and $\textbf{F}_{1}$ represents the first order linearisation effect. As indicated earlier, for small $\gamma$, we speculate that $\textbf{F}_{1}$ is proportional to the drag force, $F_{d} = 3\pi\mu a U $:
\begin{equation} \label{EQ}
 \textbf{F}_{1}  =  \textbf{\textit{g}}(\Rey,a/W, x/W, z/W) \times F_{d}
 \end{equation}
where \textbf{\textit{g}}$(\Rey,a/W, x/W, z/W)$ encodes the spatial variation across the cross-section. In this work, we do not try to analytically identify the form of the scaling function \textbf{\textit{g}}, but explore how it can be constructed by using a minimal set of full physics simulations. In fact, we show below that \textbf{\textit{g}}$(\Rey,a/W, x/W, z/W)$ (and hence, $\textbf{F}_{1}$) can be reliably constructed using just two full physics simulations. We rewrite Equation~\ref{EQ9} (for a fixed $\Rey$ and $a/W$) to compute $\textbf{F}_{1}$ as:
\begin{equation} \label{EQ10}
 \textbf{F}_{1}(\Rey, a/W, x/W, z/W)  = \frac{\textbf{F}(\Rey, a/W, \gamma, x/W, z/W) -  \textbf{F}_{0}(\Rey, a/W, x/W, z/W)}{\gamma}   
 \end{equation}

\begin{figure}
\centering
\centerline{\includegraphics{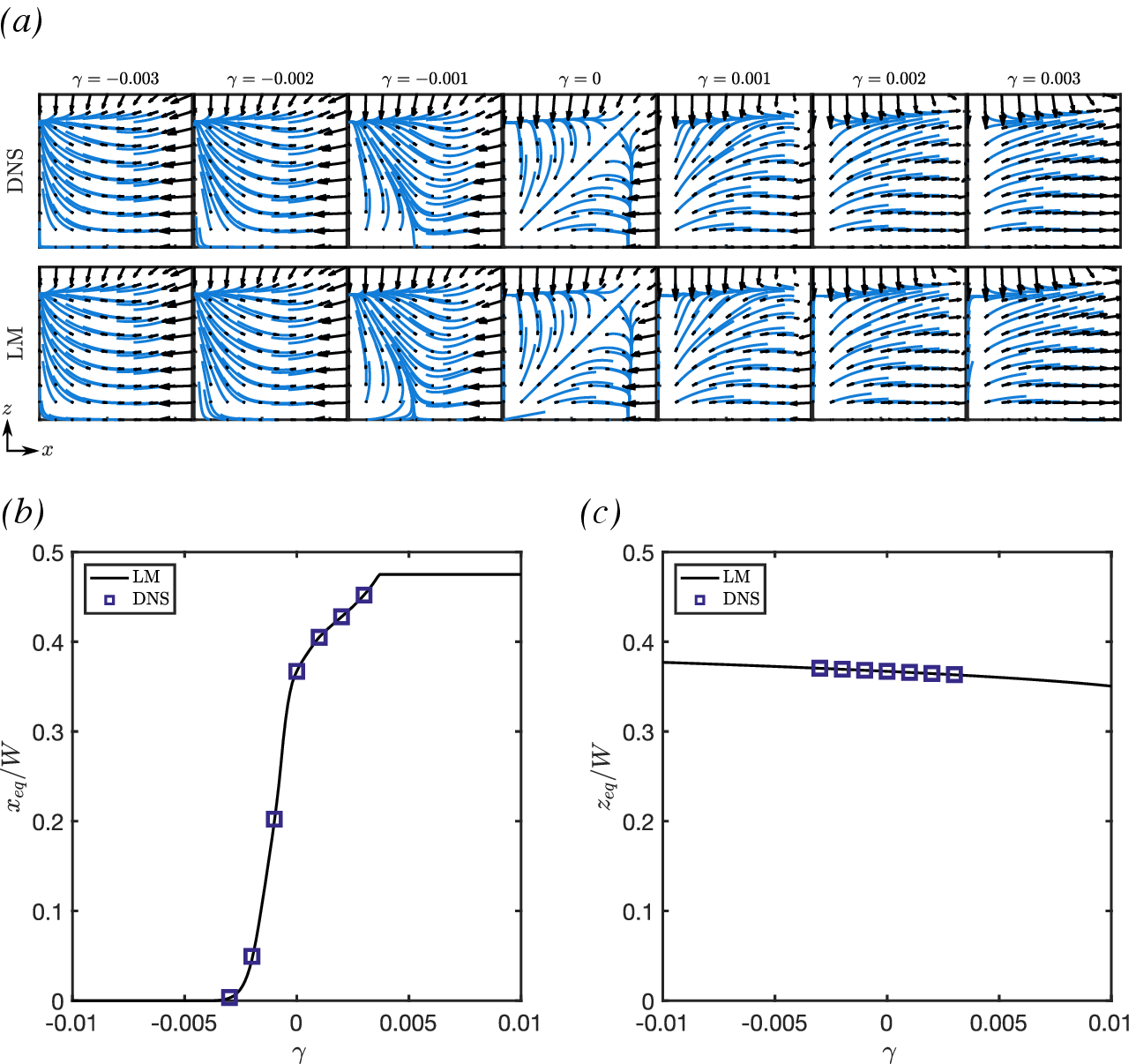}}
\caption{(\textit{a}) Force fields for various values of $\gamma$ for both the direct numerical simulation (DNS) and the linear model (LM) with a streamline trace overlaid in blue for visualisation ($a/W = 0.05$ and $\Rey = 100$). The bounding box for each field is $0.45W \times 0.45W$ (\textit{b}) $x$-equilibrium plotted against the relative permeate flow $\gamma$. (\textit{c}) $z$-equilibrium plotted against the relative permeate flow $\gamma$. The discrete points represent the results from the DNS and the continuous black line is the result of the LM.}
\label{fig:F5}
\end{figure}

Here $\textbf{F} -  \textbf{F}_{0}$ is the difference between the calculated forces maps. \textbf{F} represents the force vector field of a particle in the presence of a permeate flow at some fixed $\Rey$ and $\gamma$ and $\textbf{F}_0$ is force vector field for the same particle at the same $\Rey$ but in the absence of permeate flow ($\gamma = 0$). Figure \ref{fig:F4}a shows the spatial distribution of this residual for various values of $\gamma$. Interestingly, it appears that these curves are self-similar when scaled by $F_P = 3 \pi \mu a \gamma U$, suggesting the utility of the proposed linear model.  In figure \ref{fig:F4}b and \ref{fig:F4}c we construct \textbf{\textit{g}} by dividing the residual $\textbf{F} -  \textbf{F}_{0}$ by the characteristic permeate force $F_P$. In figure \ref{fig:F4}b, keeping $\Rey$ constant, we see three distinct curves representing our three particle sizes, showing that all normalised residual curves fall on top of each other when divided by $\gamma$. Similarly, the curves collapse as shown in figure \ref{fig:F4}c corresponding to each value of $\Rey$ (at a fixed $a/W$). We note that \textbf{\textit{g}} is insensitive to the choice of $\Rey$ (appendix \ref{App_B}).

We compare the force maps constructed using $\textbf{F}_{1}$ (\textit{i.e.} LM, for linearised model), with that from the full physics simulations (\textit{i.e.} DNS) in figure \ref{fig:F5}a. We chose the interesting case of $\Rey = 100$ and $a/W = 0.05$, and seven discrete values of $\gamma$, where fields yield drastically different `streamline' morphologies, from a complete wall suction at $\gamma = 0.003$ to centre-plane focusing at $\gamma < -0.003$. 

First we compute $\textbf{F}_{1}$  according to equation 3.3 for each of the six permeate solutions [\textit{i.e.} $\gamma = \pm 0.001, \pm 0.002, \pm 0.003$]. With this result, we average all six $\textbf{F}_{1}$ fields to produce a master solution, and thus eliminate any numerical noise that is inherent in this type of model construction. 
From figure \ref{fig:F5}, it is apparent that the LM reconstructs the lateral lift force maps qualitatively well with little discernible error; where the advantage of the LM is that it only requires knowledge of a finite number of simulations. More importantly, the LM is not limited to discrete values of $\gamma$, but can be used to evaluate forces for continuous values of $\gamma$.
We next compare the predicted location of the focussing points (force equilibrium points) in figure \ref{fig:F5}b and \ref{fig:F5}c for different $\gamma$. In figure \ref{fig:F5}b we see that the $x$-equilibrium diagram spans a wide range of the channel where the limits are set by the centreline ($x/W = 0$) and by contact with the porous wall ($x/W = 0.475$). Similarly, figure \ref{fig:F5}c shows the $z$-equilibrium diagram for the same system, where the equilibrium shift is seemingly independent of $\gamma$. The LM provides the ability to tune the permeate flow rate, $\gamma$ to precisely select a desired equilibrium location. This is potentially very useful for particle separation applications. 

  \begin{figure}
\centering
\centerline{\includegraphics{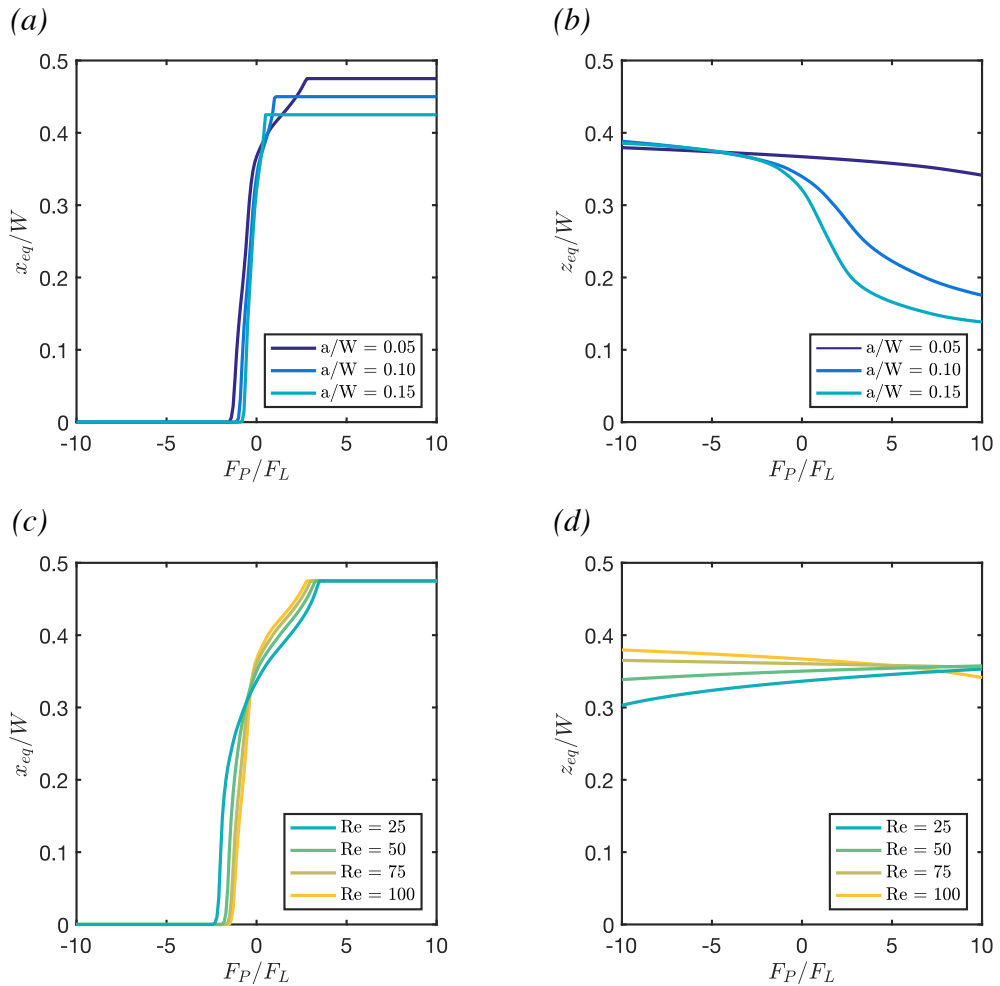}}
\caption{ (\textit{a}) $x$-equilibrium as a function of the relative permeate force ($F_P/F_L$) for three particle diametres ($a/W = 0.05,0.10$ and $0.15$) at $\Rey = 100$. Of note is that the data is limited by either the centreline ($x/W = 0$) or the confining walls for large values of ($x/W = 0.50$) \textit{b}) $z$-equilibrium diagram for three particle diametres ($a/W = 0.05,0.10$ and $0.15$). Here the equilibrium shift is less sensitive than the $x_{eq}$ counterpart that is the particle deviate only slightly from the zero permeate equilibrium. (\textit{c}) $x$-equilibrium as a function of the relative permeate force ($F_P/F_L$) for four $\Rey$ ($\Rey = 25,50,75$ and $100$) for a single particle of diametre $a/W = 0.05$. (\textit{d}) $z$-equilibrium as a function of the relative permeate force ($F_P/F_L$) for four $\Rey$ ($\Rey = 25,50,75$ and $100$) for a single particle of diametre $a/W = 0.05$.}
\label{fig:F6}
\end{figure}

We next investigate the scaling interplay between the inertial force $\textbf{F}_{0}$ and the permeate drag $\gamma\textbf{F}_{1}$ with increasing $\Rey$ and particle size, we note that  $\gamma\textbf{F}_{1}$ scales as the Stokes drag force \textit{i.e.}, $F_P \sim 3\pi \mu a \gamma U$ and the inertial lift forces scale as $F_L \sim \rho U^2 a^4/W^2$. Hence, when $|F_P| \gg F_L$ we would expect that permeate forces are dominant and the force fields would appear as if inertial forces were negligible. Conversely, when $|F_P| \ll F_L$ we would expect that inertial forces would dictate the migration and focusing of a particle in the channel. In figure \ref{fig:F6} we plot the equilibrium location against the relative permeate force $F_P/F_L$ for all particle size (figure \ref{fig:F6}a and \ref{fig:F6}b) and $\Rey$ (figure \ref{fig:F6}c and \ref{fig:F6}d). The continuous diagrams seen in figure \ref{fig:F6} show very clear and specific trends that would be difficult to interpret from discrete DNS data such as that seen in figure \ref{fig:F3}. The equilibrium diagrams shown in figures \ref{fig:F6}a and \ref{fig:F6}c clearly demonstrate the behaviour of a particle at the two extremes as predicted. That is, for large values of magnitude of $|F_P| \gg F_L$ we see a dominance of the permeate flow and the force equilibrium is shifted either to the walls ($x_{eq}/W = (W-a)/2W $) or to the centreline ($x_{eq} = 0$) and on the other extreme, when $|F_P| \ll F_L$, we see that the force equilibrium is only slightly perturbed from the case of no permeate flow. Finally, figures \ref{fig:F6}b and \ref{fig:F6}d show the change in $z$-equilibrium location as a function of the relative permeate force ({$F_P/F_L$}) for $a/W$ and $\Rey$. The $z_{eq}$ of large particles are more affected when compared to small particles under this representation of ${F_P}/{F_L}$ (figure 6b). Further, in figure 6d we see that as $F_P/F_L$ increases, the $z_{eq}$  migrates towards the channel wall for low $\Rey$ and towards the centreline for high $\Rey$. This behaviour is indicative of secondary effects that can not be captured by such a simple scaling argument \citep{DiCarlo:2009ee}.


\subsection{Limits of the linear model}
\begin{figure}
\centering
\centerline{\includegraphics{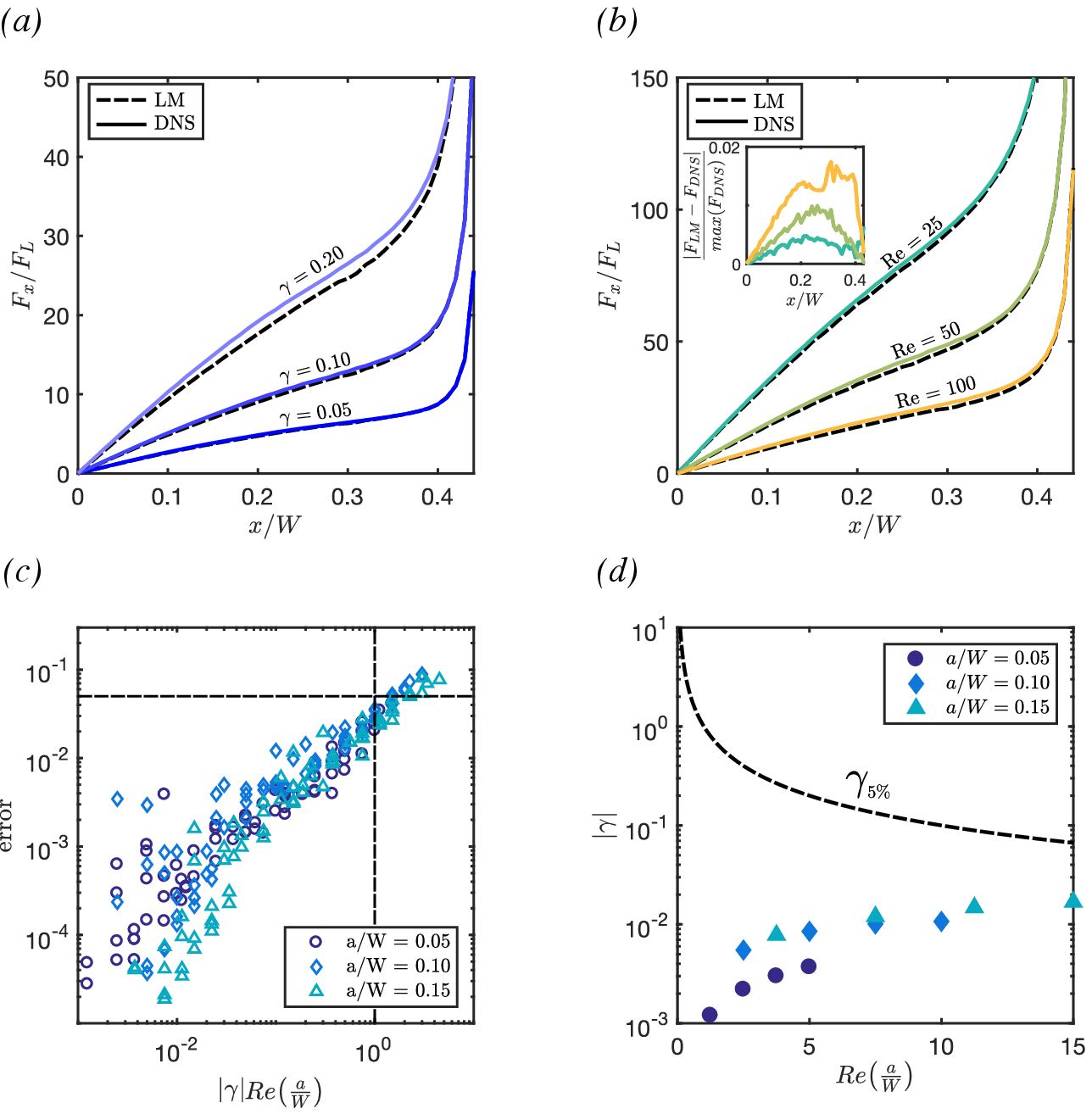}}
\caption{ A direct comparison of the forces on a particle as computed by a DNS and the LM showing how error in the LM increases with increasing $\gamma$ (\textit{a}) and $\Rey$ (\textit{b}). It is difficult to see from a direct comparisson that the error is increasing with $\Rey$ therefore, we plot in the inset the local relative error. (\textit{c}) A plot illustrating how error in the LM increases with the permeate Reynolds number $\Rey_{U_W} = |\gamma| \Rey \frac{a}{W}$. The dashed lines shows that for $\Rey_{U_W} > 1$ error in the linear model is greater than 5\%. (\textit{d}) $\gamma - \Rey \frac{a}{W}$ operating parameter space for the LM, where using the LM in the space beneath the blacked dashed line should yield less than 5\% error in the model. Superposed onto this operating parameter space are data points for the three particle sizes representing the value of $|\gamma|$ at which complete wall suction occurs for the particles studied.}
\label{fig:F7}
\end{figure}

We next explore when the additive decomposition of the force into an inertial component and a linear viscous component breaks down. We examine the error in the LM model by comparing the lateral forces of the direct numerical simulation with those constructed with our linear model to determine when nonlinear effects due to $\gamma$, $\Rey$ and $a/W$ cause the model to become unreliable. In figure \ref{fig:F7}a we show a comparison of the inertial force for three distinct $\gamma$ at $\Rey = 100$ and $a/W = 0.10$ and for three distinct $\Rey$ for $\gamma = 0.2$ and $a/W = 0.10$ (right). As expected, at increasing values of both  $\gamma$ and $\Rey$, mismatch between the two models increases. However, the increasing mismatch associated with a change in $\Rey$ is difficult to discern and so we plot in the inset to figure 7b the relative spatial error for each case.
We define the error in the LM as: 

 \begin{equation}\label{EQ11}
 error  = \frac{ \lVert  \textbf{F}_{DNS} - \textbf{F}_{LM}  \lVert_2  }      {\lVert \textbf{F}_{DNS} \lVert_2 } 
 \end{equation}

Where $\textbf{F}_{DNS}$ and $\textbf{F}_{LM}$ are the force distribution calculated using the direct numerical simulation and the linear model respectively. Figure \ref{fig:F7}b shows how the error in the LM increases with $a/W$, $\gamma$ and $\Rey$. In general, as the permeate Reynolds number $\Rey_{U_W}$ ($\Rey_{U_W} =  |\gamma| \frac{a}{W} \Rey$) increases, so does error. This confirms our hypothesis that as both inertial and permeate flow increase, nonlinear effects become more prominent and cannot be captured in our linear model.  

Over all sets of data for $\Rey_{U_W} \leq 1$ the error in the LM is less than 5\%, represented by the dashed lines in figure  \ref{fig:F7}b. Therefore, in figure \ref{fig:F7}c we determine the maximum $|\gamma|$ for a given $\frac{a}{W} \Rey$ defined as where error will be 5\%. Superimposed on this figure we show the $\gamma$ (at a given $\frac{a}{W}\Rey$) at which the particle equilibrium location coincides with the confining walls (\textit{i.e. }, wall contact) and therefore an increase in $\gamma$ will no longer result in a change in particle equilibrium. It is clear that for the particles studied in this work, that we will never reach the 5\% error limit of the LM, suggesting that the LM is a viable, useful and applicable model for fast experimental exploration of the effect inertial migration under permeate flow conditions. 

 \section{Experiments}

\begin{figure}
\centering
\centerline{\includegraphics{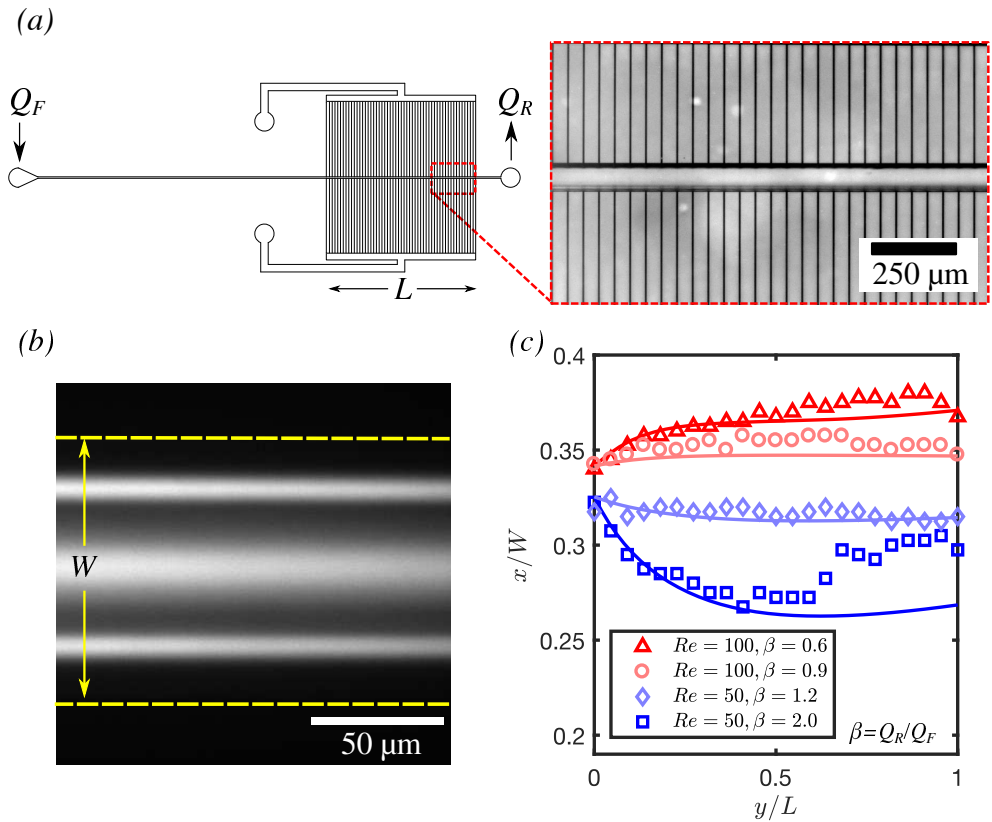}}
\caption{(\textit{a}) schematic showing the microfluidic device composed of a long straight region ($\approx$ 1.9\,cm) followed by a region ($L =  1.0\,$cm) where there is an array of perpendicular permeate channels (inset) that allow permeate flow to enter or exit the channel. (\textit{b}) Long exposure image of 10\, $\mu$m fluorescent polystyrene particles in a 100$\times$100\, $\mu$m ($W\times W$) channel. The in-plane particles are measured at a distance $x$ relative to the centreline. (\textit{c}) Comparison of computed (solid lines) and experimentally measured trajectories (shapes) of a particle $a/W = 0.10$ for various operating conditions of the microfluidic device.}
\label{fig:F8}
\end{figure}

In this section, we illustrate the utility of the linearised model by comparing experiments with predicted trajectories of inertial particles within a porous channel. We microfabricated a model porous system as seen in figure \ref{fig:F8}a. The microfluidic device is composed of a primary channel that is 3.2\,cm long with a square cross-sectional area of 100$\times$100\,$\mu$m. The particles enter the channel through a long straight region intended to prefocuss the particles. Following this region, there is a permeate region $L = 1.0\,$cm where flow can enter and exit the channel through an array of permeate channels of $W_P = $ 5\,$\mu$m width spaced $\delta =$ 50\,$\mu$m apart, with a length of $L_P = $ 4.95\,mm.  A two-syringe pump system (Harvard Apparatus) provides both inlet flow and permeate flow: the first pump infuses the inlet flow into the primary channel at a constant volumetric flow rate ($Q_F$), and the second pump has two possible configurations depending on the ratio $\beta = Q_R / Q_F$. If $\beta$ is less than one the second pump is placed at the exit of the primary channel and limits the flow rate to a rate of $Q_R$ through the primary outlet. If $\beta$ is greater than one the second pump infuses flow through the permeate channels at a rate $Q_P$ where $Q_F+ Q_P = Q_R$. In general the flow in a device like this is highly dependent on the relative hydrodynamic resistance of the permeate channels to that of the main channel. In our geometry, the permeate resistance is large compared to the main channel and it can be shown that in this limit the volumetric flow rate decreases linearly with axial location, so $Q(y) = Q_{F} [1 + \frac{y}{L}(\beta-1)]$ (appendix C).  Using conservation of mass we find the constant permeate velocity to be $U_{W} = \frac{Q_{F}(1-\beta)}{2 WL}$, and thus $\gamma(y) = W^2 U_{W}/Q(y)$. 

We performed migration experiments with a suspension of 10\,$\mu$m fluorescent polystyrene particles in DI water at a concentration of $10^{4}$ particles/ml, and we add 0.5$\%$ v/v Tween 20 (Sigma-Aldrich) to reduce particle aggregation. To find the particle locations, we record streak images with a CCD camera (Andor Luca) by accumulating approximately 25\,s of image data at each downstream location and the post-processing using image-processing software (MATLAB). Figure \ref{fig:F8}b shows an example of a streak image. We post process images like this to locate three peaks, the centre peak corresponds to the out-of-plane equilibrium while the remaining two correspond to the in-plane equilibrium. Since the in-plane equilibria are symmetric about the centreline, we build a trajectory from one in-plane equlibria at various axial locations along the length of the channel to compare with our LM. More details on the experiments can be found in \citep{Garcia:2017ho}. 

Figure \ref{fig:F8}c shows four experimental trajectories of 10$\,\mu$m particles. The prefocussed particles enter the permeate region at $y/L = 0$ and begin to migrate either towards the wall or away, depending on the direction of the permeate flow. Here values of $\beta > 1$ indicate the permeate flow is directed towards the centre of the channel and conversely $\beta < 1$ indicates that the flow is directed out of the channel. Even though the permeate flow is constant, the trajectories can be non-monotonic, where the particles begin migrating in one direction and subsequently reverse directions due to the evolving nature of inertial and permeate forces in the spatially varying flow field (figure \ref{fig:F8}c). We can calculate the particle trajectories with our linear model by knowing how the flow parameters (\textit{i.e. }$\gamma$ and $\Rey$) change with axial location in the channel. 

We calculated the theoretical trajectories with a first order time stepping approximation (\textit{i.e. }Euler method), evaluating the force at $z = 0$. We do not expect much motion in the $z$-direction because the lift forces in the $z$-direction are much smaller than the $x$-direction lift force for the prefocussed stream of particles we are modeling. Thus the equations of motion become:

\begin{equation} \label{EQ12}
 y_{n+1} = y_{n} +  u_{y}(x_{n}, z=0)\frac{Q(y)}{Q_F}\Delta t 
   \end{equation}
   
 \begin{equation}   \label{EQ13}
 x_{n+1} = x_{n} +  \frac{F_{x}(x_{n},y_n, z=0)}{3\pi \mu a \lambda} \Delta t 
  \end{equation}
 
where $u_{y}(x,z)$ is the flow field in a square channel \citep{White:2006ug} and $F_{x}(x,y,z)$ is the predicted force in the lateral direction calculated using our linear model. The axial dependence of the lateral force $F_x$ is calculated by mapping the axial location of the particle to a corresponding local value of $\Rey$. With this information, we interpolate between a pre-calculated dataset and generate the local zero permeate lift force. Using the observation that g$_x$ is invariant to $\Rey$ (figure 4c), we construct the linear model using equation 3.1. Finally, $\lambda$ is the correction factor for Stokes' drag near a confining wall (appendix B).
Using the linear model we are able to reproduce the experimental trajectories reasonably well (figure \ref{fig:F8}c), further proving the viability of our linear model.
 
\section{Conclusion}
Our findings describe the spatially inhomogeneous forces on confined inertial particles in the presence of a permeate flow. Our numerical simulations suggest that the relative permeate force ($F_P/F_L$) is an important parameter in the characterisation of behaviour of these particles. For very small magnitudes of the relative permeate force the location of force equilibrium remains unchanged and reminiscent of flow in a nonporous duct. As the magnitude of the relative permeate force is increased, the equilibrium position deviates further from the non-porous case until it is either limited by the wall or centreline. Using the results from our numerical simulations we are able to construct a model which superposes the linear viscous effects of the permeate flow to that of the underlying inertial forces. This linear model shows excellent agreement over a continuous span of permeate flow with both full simulations and experimental observations when $\Rey_{U_W} < 1$ with no added computational penalty. This is especially noteworthy because the flow field in a channel with permeable walls is not trivial and would normally require much effort to simulate with another approach. We speculate that this model can help rapidly design microfluidic devices that can precisely manipulate particle streams. Furthermore, the framework for our linear model presented in this work can also be implemented in other systems where external forcing of inertial particles exists such as with magnetic or electric forces. Our model greatly reduces the complexity of a well studied and ubiquitous flow. We emphasize the fact that the linearization, while useful, does not account for potentially important phenomena like particle deformability and non-Newtonian suspending fluids, which are typical of real world systems. Developing similar simple and computationally efficient methods for accounting for such interactions is a future area of research work.

\section{Acknowledgements}
We would like to acknowledge Professor Bassam Bamieh for enlightening discussions on nonlinear bifurcating systems. Mike Garcia was supported by the Institute for Collaborative Biotechnologies through Grants No. W911NF-09-D-0001 and No. W911NF-12-1-0031 through the U.S. Army Research Office. The content of the information does not necessarily reflect the position of the policy of the government, and no official endorsement should be inferred.


\appendix
\section{Additional momentum term}\label{App_A}

The momentum equation can be transformed from a velocity field in the lab frame to that of a frame translating with a particle moving at a velocity $U_p$ using the following transformations. 

\begin{equation} 
t' = t
\end{equation}
\begin{equation} 
x_i' = x_i + x_{p,i}(t)
\end{equation}
\begin{equation} 
u_i'= u_i + U_{p,i}(t)
\end{equation}
\begin{equation} 
p' = p
\end{equation}

Here the variable associated with a prime denote the lab frame and those without a prime denote the moving frame. We can then relate the derivatives with respect to time and space for either frame by: 

\begin{equation} 
\frac{\partial ( )}{\partial x_j} = \frac{\partial x_k'}{\partial x_j}\frac{\partial ( )}{\partial x_k'} = \frac{\partial{ (x_k+x_{p,k})}}{\partial x_j} \frac{\partial ()}{\partial x_k'}= \delta_{jk} \frac{\partial ()}{\partial x_k'} = \frac{\partial ( )}{\partial x_j'}
\end{equation}

\begin{equation} 
\frac{\partial ( )}{\partial t} = \frac{\partial t'}{\partial t} \frac{\partial ()}{\partial t'} + \frac{\partial x_k'}{\partial t}\frac{\partial ( )}{\partial x_k'} = \frac{\partial ()}{\partial t'} + U_{p,k} \frac{\partial()}{\partial x_k'}
\end{equation}

We can then relate the momentum equation in the lab frame to that of a moving frame by using equation A5 \& A6

\begin{equation} 
\rho \bigg[ \frac{\partial u_i'}{\partial t'} + u_j' \frac{\partial u_i'}{\partial x_j'}\bigg]  = - \frac{\partial p'}{\partial x_i'} + \mu \frac{\partial}{\partial x_j'}\bigg( \frac{\partial u_i'}{\partial x_j'} \bigg)
\end{equation}

\begin{equation} 
\rho \bigg[ \frac{\partial (u_i + U_{p,i})}{\partial t'} + (u_j+U_{p,j}) \frac{\partial (u_i+U_{p,i})}{\partial x_j'}\bigg]  = - \frac{\partial p}{\partial x_i'} + \mu \frac{\partial}{\partial x_j'}\bigg( \frac{\partial (u_i+U_{p,i})}{\partial x_j'} \bigg)
\end{equation}

\begin{equation} 
\rho \bigg[\frac{\partial (u_i + U_{p,i})}{\partial t} - U_{p,j} \frac{\partial(u_i + U_{p,i})}{\partial x_j} + (u_j+U_{p,j}) \frac{\partial (u_i+U_{p,i})}{\partial x_j}\bigg]  = - \frac{\partial p}{\partial x_i} + \mu \frac{\partial}{\partial x_j}\bigg( \frac{\partial (u_i+U_{p,i})}{\partial x_j} \bigg)
\end{equation}

Now canceling like terms and knowing that $U_{p,i}$ has no spatial gradients because it is only a linear translation of the laboratory frame and not a continuum value, we can write the momentum equation as: 

\begin{equation} 
\rho \bigg[\frac{\partial (u_i + U_{p,i})}{\partial t}  + u_j \frac{\partial u_i}{\partial x_j}\bigg]  = - \frac{\partial p}{\partial x_i} + \mu \frac{\partial}{\partial x_j}\bigg( \frac{\partial u_i}{\partial x_j} \bigg)
\end{equation}

We assume that in the moving reference frame the flow is quasi-steady and thus the time derivative of the flow is zero, yielding: 

\begin{equation} 
\rho \bigg[\frac{\partial U_{p,i}}{\partial t}  + u_j \frac{\partial u_i}{\partial x_j}\bigg]  = - \frac{\partial p}{\partial x_i} + \mu \frac{\partial}{\partial x_j}\bigg( \frac{\partial u_i}{\partial x_j} \bigg)
\end{equation}

At a given moment in time the acceleration of the particle that is being tracked by the moving reference frame is dictated by the underlying flow ($\bar{u}_i$) and its gradient and thus the acceleration is given by

\begin{equation} 
\frac{\partial U_{p,i}}{\partial t} = U_{p,j} \frac{\partial \bar{u}_i}{\partial x_j} 
\end{equation}

Finally, substituting this term back into the momentum equation yields:

\begin{equation} 
\rho \bigg[U_{p,j} \frac{\partial \bar{u}_i}{\partial x_j}   + u_j \frac{\partial u_i}{\partial x_j}\bigg]  = - \frac{\partial p}{\partial x_i} + \mu \frac{\partial}{\partial x_j}\bigg( \frac{\partial u_i}{\partial x_j} \bigg)
\end{equation}

where $\bar{u}$ is the undisturbed velocity field. However, the only non-zero component of (A12) in our model acts in the axial direction and scales with $\gamma$ as:

\begin{equation} 
\frac{\partial U_{p,i}}{\partial t} = U_{p,j} \frac{\partial \bar{u}_i}{\partial x_j} \sim U_{p}\frac{\gamma U}{W}
\end{equation}

Note the term is negligible for small $\gamma$ and near the channel walls where $U_p$ approaches zero. 


\section{Correction to Stokes' drag and the relationship to \textit{g}}\label{App_B}

When the size of a flowing particle is comparable to the dimensions of its confining channel, it is important to consider the hydrodynamic effects from the walls on the lateral motion of the particle. To account for this, we introduce a spatially varying retardation factor $\lambda$ that relates force ($F$) on a particle of diametre($a$) and its migration velocity ($U$) (Brenner 1961):  

\begin{equation} 
F = 3 \pi \mu a U \lambda
\end{equation}

Here, $\lambda$ is calculated with a numerical simulation where a particle is modeled in a square channel of cross-section $W \times W$ and is assigned a velocity $U$ ($\Rey \ll 1$) in the direction of a wall. Then we calculate the required force to produce such a motion by integrating the fluid stress on the surface of the particle and using equation B1 we solve for $\lambda$. We repeat this process for discrete locations that span the width of the channel. In figure 9a we plot $\lambda$ for three particles ($a/W = 0.05, 0.10, 0.15$ at $z/W = 0$) and show how our calculations compare to the existing analytic model \citep{Brenner:1961eb}. Interestingly, even though in our simulations the particle is confined by four walls, the semi-infinite domain of the Brenner model seems to match well at $z/W = 0$. 
In our work we find that the normalised residual \textbf{\textit{g}} (figure \ref{fig:F4} b \& c) is weakly dependent on $\Rey$. A quick approximation of \textbf{\textit{g}} (for any sized particle) may useful for any researcher who may be interested in applying this linear model. Here we provide a simple method for approximating \textbf{\textit{g}}.
We begin by interpreting \textbf{\textit{g}} to represent the effects of the permeate flow ($U_W$) on the force ($\textbf{F}_P$) experienced by a particle. 

\begin{equation}  
\textbf{F}_P = \gamma \textbf{F}_1 = 3 \pi \mu a U_W \textbf{\textit{g}}
\end{equation}

Likewise, if we interpret Equation B1 as the force experienced by particle with a flow moving past it (\textit{i.e.} particle reference frame) then we can construct a force field everywhere in the channel according to the local permeate velocity given by the $x$, and $z$ components of undisturbed flow $\bar{\textbf{u}}$ that should approximate the permeate force field.

\begin{equation}  
\textbf{F}_p \approx  3 \pi \mu a \bar{\textbf{u}} \lambda
\end{equation}

Without loss of generality we consider only the $x$-component of the permeate force. If we equate equation B2 and B3 we can then use this relationship to obtain $g_x$

\begin{equation}  
g_x \approx \lambda \frac{\bar{u}_x}{U_W}
\end{equation}

Here $\bar{u}_x $ is $x$-component of the undisturbed flow and can be seen in figure 9b. The results of implementing equation B4 can be seen in figures 9 c-e. Where in general the approximation works best for smaller particles at a lower $\Rey$. However, the approximation is remarkably accurate and only requires knowledge of flow field, which is relatively simple to determine.

\begin{figure}
\centering
\centerline{\includegraphics{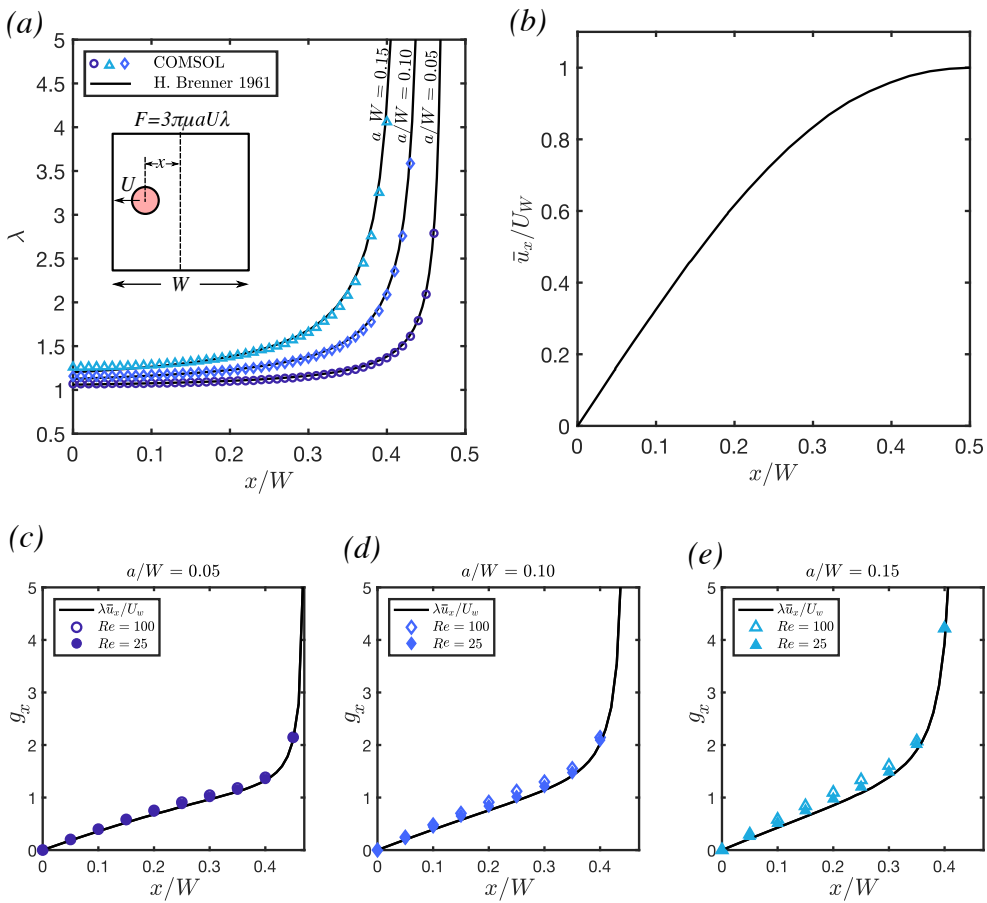}}
\caption{
\textit{(a)}
The relationship between the translational velocity of a sphere ($U$) and the force ($F$) required to produce such a motion can be significantly modified by the effects of confining walls. We show the effect of this spatial retardation  ($\lambda$) on spheres of diametre $a/W = 0.05,0.10,0.15$ (at $z/W = 0$) and compare with the analytic results of \citep{Brenner:1961eb}.
\textit{(b)}
A plot of the spatially varying permeate velocity (at $z/W = 0$) within the channel modeled in this work. At the walls the permeate velocity is maximal ($u_x/U_W = 1$) and decays to zero at the centre of the channel.
\textit{(c)}
Normalised residual curves ($g_x$) as a function of $x/W$ at $z/W =0$ for a particle of diametre $a/W = 0.05$
\textit{(d)}
Normalised residual curves ($g_x$) as a function of $x/W$ at $z/W =0$ for a particle of diametre $a/W = 0.10$
\textit{(e)}
Normalised residual curves ($g_x$) as a function of $x/W$ at $z/W =0$ for a particle of diametre $a/W = 0.15$
The solid black lines in \textit{(c)-(e)} represent $g_x$ modeled using the retardation factor and permeate velocity depicted in \textit{(a)} and \textit{(b)} respectively.}
\label{fig:F10}
\end{figure}


\section{Local volumetric flow rate in a porous channel}\label{App_C}
The local volumetric flow rate in a porous duct $Q(y)$ is typically axially varying due to a non-zero fluid flux. Thus the local volumetric flow rate can be modeled as: 

\begin{equation} \label{eq:B1}
 Q(y) = Q_F - \int_{0}^{y} U_W(y')\,Wdy'
\end{equation}

Where $Q_F$ is the feed flow rate \textit{i.e.}, $Q(y=0) = Q_F$, $U_W$ is the average permeate flow velocity, $W$ is the channel height and $y$ is the axial coordinate. Now assuming that the wall can be treated as a continuously porous wall of a constant permeability we can use Darcy's law to write. 

 \begin{equation} \label{eq:B2}
 Q(y) = Q_F - \int_{0}^{y} \kappa \frac{P(y')-P_0}{\mu L_P} \,W dy'
\end{equation}

Where $P(y')$ is the local pressure in the channel, $P_0$ is the reservoir pressure on the other side of porous wall, $L_p$ is the wall thickness and $\kappa $ is the wall permeability. It is convenient to define the constant  $R_P =\mu L_P/  \kappa W$ so that we can rewrite equation \ref{eq:B2} 

\begin{figure}
\centering
\centerline{\includegraphics{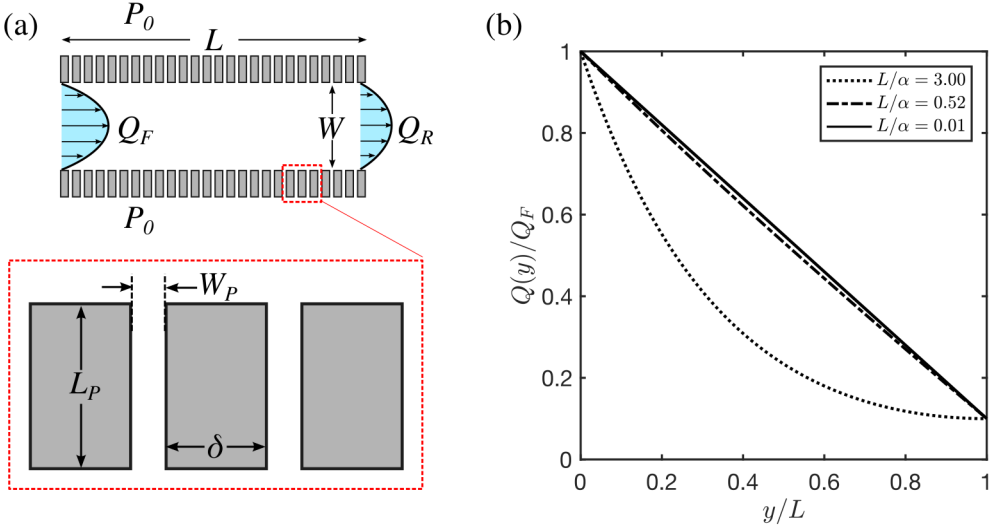}}
\caption{(\textit{a}) schematic showing the microfluidic device used in the experiments. The channel has a height W into the page  (\textit{b}) volumetric flow rate curves for channels of varying channel resistance $L/\alpha$ ($\alpha^2 = R_P/R_C$) for a $\beta = 0.1$ ($\beta = Q_R/Q_F$) as $L/\alpha$ approaches 0 the axial flowrate distribution becomes linear.}
\label{fig:F10}
\end{figure} 

 \begin{equation}  \label{eq:B3}
 Q(y) = Q_F - \int_{0}^{y}  \frac{P(y')-P_0}{R_P} \,dy'
\end{equation}

Equation \ref{eq:B3} can be rewritten in terms of pressure by using the Hagen-Poiseuille equation:
  
 \begin{equation} \label{eq:B4}
Q = -\frac{1}{R_C} \frac{dP}{dy}
\end{equation}

Here $R_C$ represents the channel resistance per unit length in the axial direction. This equation can now be substituted into \ref{eq:B3}

 \begin{equation}  \label{eq:B5}
-\frac{1}{R_C} \frac{dP}{dy} = Q_F - \int_{0}^{y}  \frac{P(y')-P_0}{R_P} \,dy'
\end{equation}

If we differentiate equation \ref{eq:B5} once and use the fundamental theorem of calculus we obtain a ordinary differential equation for the axial pressure distribution.

 \begin{equation}  \label{eq:B6}
-\frac{1}{R_C} \frac{d^2P}{dy^2} =  - \frac{P(y)-P_0}{R_P} 
\end{equation}

Now rearranging the equation to a standard form

\begin{equation}  \label{eq:B7}
\frac{R_P}{R_C} \frac{d^2P}{dy^2} - P(y) + P_0 = 0
\end{equation}

say that $\alpha^2 = R_P/R_C$ and solve the ODE

\begin{equation}  \label{eq:B8}
P(y) = C_1 \exp{\left(\frac{y}{\alpha} \right)} + C_2 \exp{\left(\frac{-y}{\alpha} \right) }+P_0
\end{equation}

This solution is valid however the known boundary conditions are in terms of volumetric flowrate. Therefore, it is convenient to differentiate the solution once 

\begin{equation}  \label{eq:B9}
\frac{dP}{dy} =  \frac{C_2}{\alpha}\exp{\left(\frac{-y}{\alpha} \right)} - \frac{C_1}{\alpha} \exp{\left(\frac{y}{\alpha} \right)} 
\end{equation}

and then again apply the Hagen-Poiseuille relationship to obtain the axial flowrate distribution 

\begin{equation}  \label{eq:B10}
Q(y) = \frac{C_1}{ \alpha R_C} \exp{\left(\frac{y}{\alpha} \right)} - \frac{C_2}{\alpha R_C}\exp{\left(\frac{-y}{\alpha} \right)}
\end{equation}

Where the boundary conditions are $Q(0) = Q_F$ and $Q(L) = Q_R$ and are used to solve for $C_1$ and $C_2$

\begin{equation}  \label{eq:B11}
C_1 = \frac{-\alpha R_C  \exp \left(\frac{L}{\alpha}\right) \left [Q_R - Q_F\exp \left(\frac{-L}{\alpha}\right)\right]}  {\exp \left(\frac{2L}{\alpha}\right) - 1}
\end{equation}

and 

\begin{equation}  \label{eq:B12}
C_2 = \frac{-\alpha R_C  \exp \left(\frac{L}{\alpha}\right) \left [Q_R - Q_F\exp \left(\frac{L}{\alpha}\right)\right]}  {\exp \left(\frac{2L}{\alpha}\right) - 1}
\end{equation}

upon simplification 

\begin{equation}  \label{eq:B13}
Q(y) = \frac{Q_R \sinh\left(\frac{y}{\alpha} \right) + Q_F \sinh\left(\frac{L-y}{\alpha} \right) }{ \sinh\left(\frac{L}{\alpha}\right)}
\end{equation}

This model is sufficient to model the flow rate in a porous channel, but it can be further simplified in the limit where the flow resistance through the channel wall is much greater than the resistance through the primary channel $L/\alpha << 1$  and $\sinh(L/\alpha) \approx L/\alpha$ and in this limit equation \ref{eq:B13} reduces down to 

\begin{equation}  \label{eq:B14}
Q(y) = Q_{F} \left[1 + \frac{y}{L}(\beta-1)\right] \quad and \quad  \beta  = Q_R/Q_F
\end{equation}

The resistance of a rectangular section of channel of height $2b$ and width $2a$ ($a>b$) is given by \citep{White:2006ug}.

\begin{equation}  \label{eq:B15}
R = \frac{3\mu}{4ba^3\left(1-\frac{192a}{\pi^5b}\sum\limits_{n=1,3,5,...}^{\infty} \frac{\tanh(n\pi b/2a)}{n^5}\right)}
\end{equation}

For a square channel of width, $W$ height, $W$ the axial channel resistance is given by 

\begin{equation}  \label{eq:B16}
R_C  \approx  \frac{28 \mu}{W^4}
\end{equation}

The permeability of a porous material is defined as 

\begin{equation}  \label{eq:B17}
\kappa =  \mu \bar U \frac{L_p} {\Delta P}
\end{equation}

choosing the right formulation for average permeate velocity $\bar U_W$ is not obvious, but one reasonable approximation is to treat the wall as a continuously permeable (rather than discretely) but then use the resistance of single permeate channel to calculate the volumetric flow rate for a section of the channel. 

\begin{equation}  \label{eq:B18}
Q =2 \bar U_W W(W_P+\delta) =  \left[1-\frac{192 W}{\pi^5 W_p}\sum\limits_{n=1,3,5,...}^{\infty} \frac{\tanh(n \pi W_p/2W)}{n^5}\right] \frac{W^3 W_P}{12 \mu} \frac{\Delta P}{L_p}
\end{equation}

Equation \ref{eq:B17} can be used to solve for an expression giving $\bar U_W$. Here $\Delta P = P(y) - P_0$ and $W_P$ is the width of the permeate channels (figure 10a). Knowing $\bar U_W$ we can substitute that into the definition of $\kappa$ (equation \ref{eq:B16}) then substitute  $\kappa$ into the definition of $R_P$ to arrive at:

\begin{equation}  \label{eq:B19}
R_P \approx  \frac{24 \mu L_P(W _p + \delta)}{W^3W_P[1-0.63(W/W_P)\tanh(\pi W_p/2W)]}
\end{equation}

If we use values that are representative of our microchannel we can calculate a value of $\alpha \approx 0.0194$\,m and the resulting ratio $L/ \alpha = 0.52$. This ratio is of $\mathcal{O}(10^{-1})$ which is sufficiently small to use the small angle approximation that results in the distribution of equation \ref{eq:B14}. Figure 1b shows that as $L/\alpha$  decreases we observe a more linear trend in the axial flowrate distribution and for $L/\alpha = 0.01$ the distribution is for all practical purposes linear which is approximated well by $L/ \alpha = 0.52$.


\bibliographystyle{jfm}
\bibliography{JFB_BIB}
\end{document}